\documentclass[twocolumn,showpacs,preprintnumbers,amsmath,amssymb,prb,floatfix]{revtex4}
\pdfoutput=1

\usepackage{graphicx,tabularx}
\usepackage{amssymb}
\usepackage{dcolumn}
\usepackage{color}
\usepackage[mathcal]{euscript}
\definecolor{gray0}{gray}{0.0}
\definecolor{gray64}{gray}{0.25}
\definecolor{gray128}{gray}{0.5}
\definecolor{gray192}{gray}{0.75}
\definecolor{gray255}{gray}{1.0}

\begin{document}
\title{Formation of Pt induced Ge atomic nanowires on Pt/Ge(001): a DFT study.}
\author{Danny E. P. Vanpoucke}
\affiliation{Computational Materials Science, Faculty of Science
and Technology and MESA+ Institute for Nanotechnology, University
of Twente, P.O. Box 217, 7500 AE Enschede, The Netherlands}
\author{Geert Brocks}
\affiliation{Computational Materials Science, Faculty of Science
and Technology and MESA+ Institute for Nanotechnology, University
of Twente, P.O. Box 217, 7500 AE Enschede, The Netherlands}
\date{\today}
\begin{abstract}
Pt deposited onto a Ge(001) surface gives rise to the spontaneous
formation of atomic nanowires on a mixed Pt-Ge surface after high
temperature annealing. We study possible structures of the mixed
surface and the nanowires by total energy (density functional
theory) calculations. Experimental scanning tunneling microscopy
images are compared to the calculated local densities of states.
On the basis of this comparison and the stability of the
structures, we conclude that the formation of nanowires is driven
by an increased concentration of Pt atoms in the Ge surface
layers. Surprisingly, the atomic nanowires consist of Ge instead
of Pt atoms.
\end{abstract}

\maketitle

Self-assembly at surfaces forms an attractive method to engineer
nanostructures.\cite{Barth:nat05} In recent years we have seen a
rapid development in techniques to grow metal atomic nanowires
(NWs) on semiconductor substrates by self-assembly. NWs have been
made by adsorption of metals on planar or vicinal Si and Ge
surfaces,
\cite{Yeom:prl99,Lee:prl05,Ahn:prl05,Snijders:prl06,Eames:prb06}
and by metal-silicide or -germanide formation at Si or Ge
surfaces.\cite{LinXF:prb93,Yeom:prl05,Gurlu:apl03,Schafer:prb06,Oncel:prl05}
From a fundamental point of view these metallic NWs show the
exotic physical properties typical of (quasi) one-dimensional
systems, such as Peierls-like instabilities, charge density waves
and L\"uttinger liquid behavior. From the perspective of
applications metal NWs offer the prospect of interconnects for
quantum- and nanodevices.

Recently, G\"url\"u \textit{et al.} produced arrays of NWs
by depositing Pt on a Ge(001) surface. Perfectly straight, defect
free and regularly spaced NWs with a cross-section of one
atom and a length of up to one micron, are formed after annealing
at $T=1050$~K.\cite{Gurlu:apl03} The structures are
studied by scanning tunneling microscopy and
spectroscopy (STM and STS),
\cite{Oncel:prl05,Houselt:nanol06,Fischer:prb07}
characterizing the electronic states around the Fermi level. However,
hampered by the lack of chemical information in STM, so far only a
tentative model for the atomic structure of the NWs
exists.\cite{Gurlu:apl03} Deposition of $\sim$0.25 monolayer (ML) Pt on Ge(001) at room
temperature creates a surface with a high amount of defects
and no clear identification of Pt atoms.\cite{Gurlu:prb04}
Subsequent annealing of this surface results in the formation of patches of two different structures, the so-called $\alpha$- and $\beta$-surfaces. It has been proposed that 0.25 ML Pt is incorporated in the top surface layer of the $\beta$-surface.\cite{Gurlu:apl03,Fischer:prb07} After the same annealing step, part of the $\beta$-surfaces are covered with NWs. On the basis of available STM data, the wires
have been tentatively identified as Pt wires.\cite{Gurlu:apl03,Fischer:prb07}

In this paper we present a computational study of the structure of
the $\beta$-surface and the NWs at the first-principles density
functional theory (DFT) level.\cite{compdetails} By calculating
total energies and comparing simulated to experimental STM images
we identify the most probable structures.\cite{simudetails} The
$\beta$-terrace has a structure that is similar to the clean
Ge(001) surface, but with one in four Ge atoms replaced by a Pt
atom. The process of the formation of NWs is driven by an increase
in the concentration of Pt in the surface layers. Most remarkably,
we predict that the NWs that are observed in STM in fact consist
of \textit{Ge atoms} that are displaced from the substrate.

\begin{figure}[!tbp]
  \includegraphics[width=7.0cm,keepaspectratio]{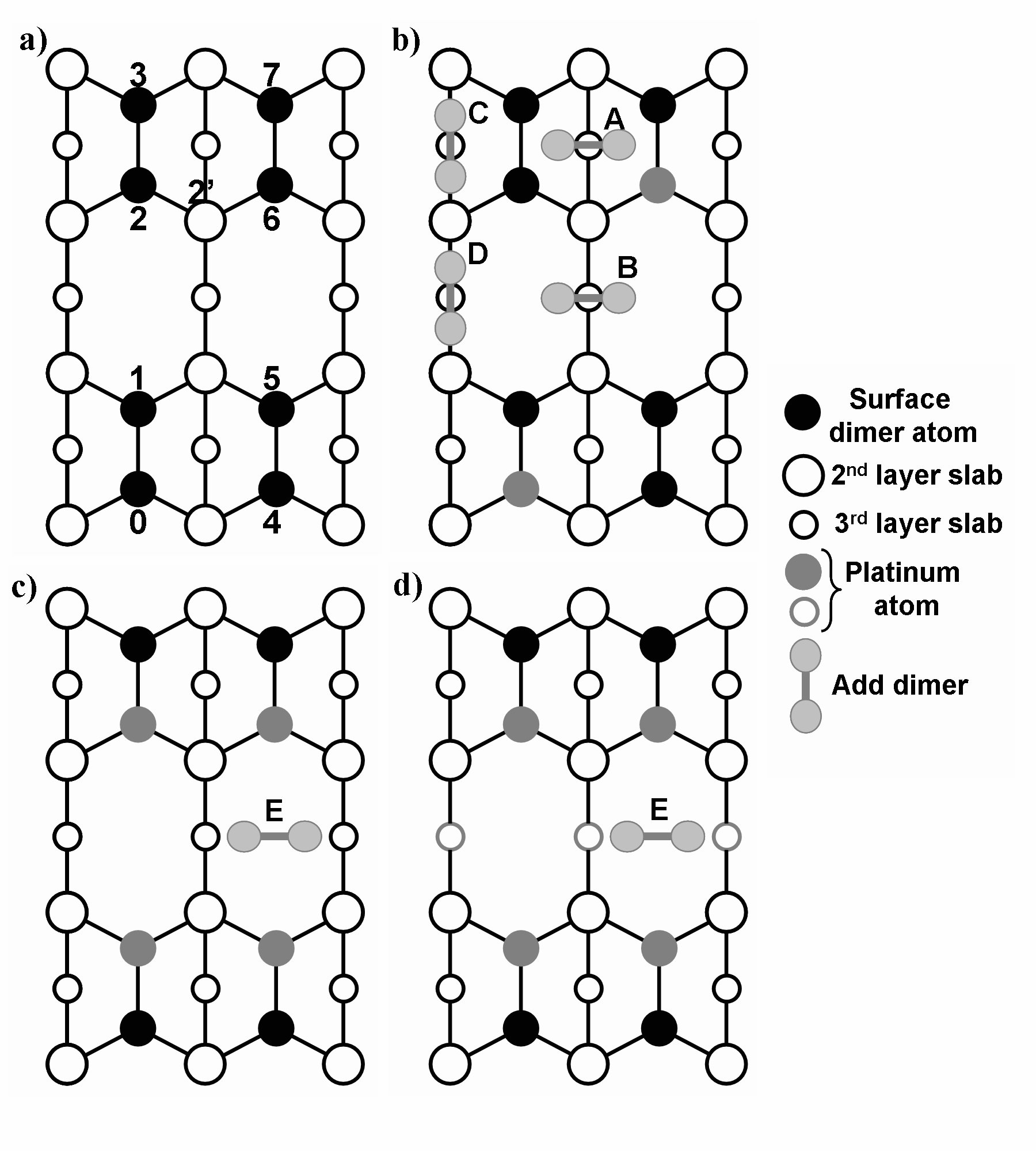}
  \caption{Schematic representation of Pt-Ge(001) surfaces.
  (a) $\beta$-surfaces have $\frac{1}{4}$ ML of Pt atoms in the top layer with
  possible positions given by the indices. (b) Adsorption sites for ad-dimers
  on the $\beta_6$-surface (Pt atoms at positions 0,6). (c) and (d) adsorption of a NW (E) on the $\gamma$-surface (Pt atoms at positions 1,2,5,6) and the $\gamma^*$-surface (additional Pt atoms in the third layer under the NW), respectively.}\label{fig1:geometries}
\end{figure}

Before discussing the structure of NWs, we study possible geometries of the $\beta$-surface. From
STM images one can conclude that the latter has a basic structure similar to that of the clean
Ge(001) surface. The top surface layer consists of rows of dimers, as shown schematically in Fig.
\ref{fig1:geometries}(a). Compared to a clean Ge(001) surface with $(1\times2)$ reconstruction,
the surface unit cell of the $\beta$-terrace is doubled, leading to a $c(2\times4)$
periodicity.\cite{Gurlu:apl03} Based upon the pattern observed in
STM, it has been proposed that the top surface layer of the $\beta$-surface contains 0.25 ML of
Pt.\cite{Gurlu:apl03}

We calculate the total energies of possible $\beta$-surface
structures by replacing one in four Ge atoms in the top surface layer of Ge(001) by Pt atoms. All
possible arrangements of Pt atoms in a $p(2\times4)$ cell are
considered, see Fig. \ref{fig1:geometries}(a). The cell
contains two Pt atoms; the first is placed at position 0, and the
second platinum atom at one of the positions 1-7. We use $\beta_n$ to indicate the structure with the second atom at position $n$. After relaxing the geometries,
surface formation energies $E_{\mathrm{f}}$, normalized per
$p(2\times4)$ unit cell, are calculated from
\begin{equation}
E_{\mathrm{f}} =
E_{\mathrm{rec}}-E_{\mathrm{Ge(001)}}-N_\mathrm{Pt}E_{\mathrm{Pt}}-\Delta{N_\mathrm{Ge}}E_{\mathrm{Ge}}.
\end{equation}
$E_{\mathrm{rec}}$ and $E_{\mathrm{Ge(001)}}$ are the total
energies of the slabs representing the surface containing Pt atoms
and the clean $p(1\times2)$ Ge(001) surface.\cite{factornote}
$N_\mathrm{Pt}$ is the number of Pt atoms and $\Delta{N_\mathrm{Ge}}$ is the
difference between the slabs in the number of Ge atoms;
$E_{\mathrm{Pt}}$, $E_{\mathrm{Ge}}$ are the energies per atom of bulk Pt and Ge. Positive/negative values of $E_{\mathrm{f}}$ indicate that the surface is unstable/stable with respect to phase separation into Ge(001) and bulk Pt. As we will see below, Pt has a tendency to be incorporated in the Ge-surface forming Pt-Ge bonds. Moreover, surfaces with low Pt density tend to be unstable with respect to phase separation into Ge(001) and surfaces with high Pt content.


The $\beta_1$ structure, with Pt in adjacent
positions (0,1), has a large positive formation energy of $\sim
0.6$ eV. It indicates that formation of Pt-Pt dimers is very
unfavorable. In contrast, several structures with mixed Pt-Ge
dimers have a negative formation energy. The most stable structure
is the $\beta_4$ geometry (Pt at positions $0$ and $4$)
having a formation energy $E_{\mathrm{f}} = -119$ meV. This
structure consists of alternating rows of Ge-Ge and Pt-Ge dimers
with all Pt atoms at the same side of a row, leading to a
$p(1\times4)$ periodicity. The second most stable structure is the
$\beta_6$ geometry (Pt at positions $0$ and $6$) with
$E_{\mathrm{f}} = -48$ meV. The $\beta_6$ structure gives a
checkerboard pattern of Pt-Ge and Ge-Ge dimers with $c(2\times4)$
periodicity.

Several $\beta_n$ geometries with mixed Pt-Ge dimers are
close in energy, which means that they are thermodynamically
accessible at the formation temperature of the $\beta$-surface (1050K).
The energy difference between the $\beta_4$ and the
$\beta_6$ structures is only $35$ meV per Pt atom, for example. Studying the formation kinetics is beyond the present calculations. To identify which of the $\beta_n$ structures may explain the experimental STM results, we calculate STM images within the Tersoff-Hamann approach.\cite{simudetails} Comparison to experimental data shows good agreement for the $\beta_6$ structure only. Other $\beta_n$ structures can be ruled out as they lead to a different
periodicity or qualitatively different STM patterns.

\begin{figure}[!tbp]
  \includegraphics[width=8.5cm,keepaspectratio]{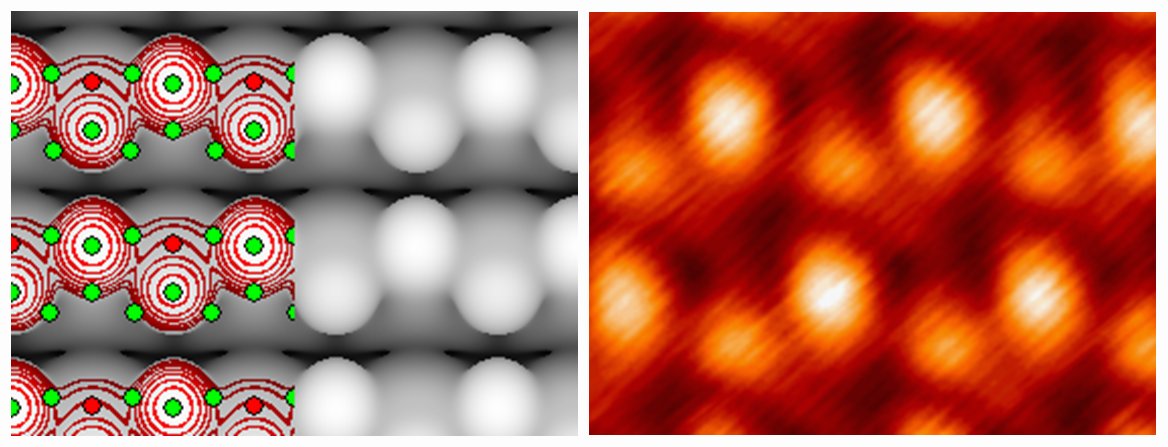}
  \caption{(Color online) Left: simulated filled state STM image of the
  $\beta_6$-surface (Pt atoms at positions 0,6) at bias $V=-0.70$ V.\cite{simudetails}
  Contours are added to guide the eye. Red(dark gray) and green(light gray) discs show
  positions of Pt and Ge atoms, respectively. Right: experimental STM image of the $\beta$-surface at $V=-0.3$ V.\cite{fn:Fig2Zand}}\label{fig2:STMs}
\end{figure}

\begin{figure}[!tbp]
  \includegraphics[width=8.5cm,keepaspectratio]{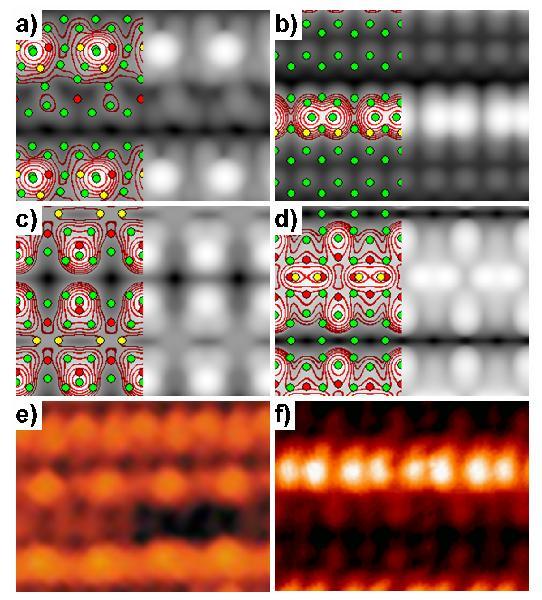}
  \caption{(Color online)
  (a)-(d) Simulated STM images at bias $V=-1.50$ V, see Fig.~\ref{fig2:STMs}.\cite{simudetails} The positions of NW adatoms are represented by yellow(white) discs. (a) The structure after optimizing a Pt NW on the $\beta_6$-surface; (b) Pt NW on Ge(001); (c) (sunken) Pt NW on the $\gamma$-surface; (d) Ge NW on the $\gamma^*$-surface, see Fig.~\ref{fig1:geometries}. (e) Experimental STM image of a wide trough; $V=-0.50$ V, $I=0.50$ nA.\cite{Fischer:prb07} (f) Experimental STM image of a NW; $V=-1.35$ V, $I=0.50$ nA.\cite{fn:Fig2Zand}}\label{fig3:STMs}
\end{figure}

Experimentally NWs are always found on patches of $\beta$-surface.\cite{Gurlu:apl03,Fischer:prb07} In addition, the NWs
are clearly composed of ad-dimers. Therefore, as a first scenario
we study the $\beta_6$ structure as a template for the adsorption
of Pt ad-dimers. Some possible geometries are sketched in Fig.
\ref{fig1:geometries}(b). Remarkably, none of the structures seem to be stable
and optimizing the geometries can lead to large
displacements of adsorbed atoms and of atoms in the
substrate. For instance, the formation energies of the optimized
structures resulting from initial adsorption of Pt dimers at
A,B,C,D sites are $E_{\mathrm{f}} = 1.78,-1.41,0.12,1.30$ eV,
respectively. Although the B structure seems to be favorable,
inspection of the optimized geometry shows that it is completely
different from the initial adsorption of a Pt dimer at a B site.
The adsorbed Pt dimer breaks up into two atoms. One Pt atom remains in the trough between the dimer rows, but sinks into the surface to form bonds with nearby Ge atoms. The second Pt atom is exchanged with the Ge atom at position 2 in the surface. The displaced Ge atom is pushed out of the surface above position $2^\prime$ (cf. Fig. \ref{fig1:geometries} (a)). The reordering of Pt and Ge atoms at the surface again indicates that the formation of Pt-Ge bonds is energetically strongly favored. The displaced Ge atom forms the highest point on the surface and is the most prominent feature in the simulated filled state STM image. The pattern however does not resemble that of a NW, compare Figs. \ref{fig3:STMs}(a) and (f).

In a second possible scenario the Pt atoms comprising a NW are
kicked out from a $\beta$-surface, whereby the latter is
transformed back into a Ge(001) surface. To investigate this
scenario we calculate the geometries and energies of Pt dimers
adsorbed on a clean Ge(001) surface. None of the structures turn
out to be stable, and often lead large atomic displacements in the
substrate. For example, the formation energies of the optimized
structures starting from the A,B,C,D configurations, are
$E_{\mathrm{f}} = 2.08,0.36,2.26,0.18$ eV, respectively. None of
the simulated STM images correspond to what is observed
experimentally. Fig. \ref{fig3:STMs}(b) shows the optimized B
configuration. Remarkably, the bright features belong to Ge atoms
that are displaced from the substrate; the Pt atoms remain
invisible.

In conclusion, the scenario's discussed in the previous two
paragraphs are unlikely. Pt adatoms do not form a NW, but instead
show a tendency to be embedded in the surface and form
additional Pt-Ge bonds. The next logical step therefore is to
consider a substrate where all dimers in the surface top layer are
Pt-Ge dimers. Structures with the Pt atoms on the same side of a
Pt-Ge dimer row are the most stable. The formation energy of the
structure shown in Fig. \ref{fig1:geometries}(c) is
$E_{\mathrm{f}} = -0.25$ eV, demonstrating that energetically this
structure is reasonable. We call this structure the
$\gamma$-surface; it consists of Pt-Ge dimers with Pt atoms at
positions 1,2,5,6. The Pt-Ge dimers form rows along the horizontal direction
in Fig. \ref{fig1:geometries}(c), not unlike the dimer
rows on the clean Ge(001) surface. There are two kinds of troughs
above the atoms in the third layer. The first kind is lined with
Pt atoms in the surface top layer and the second kind with Ge
atoms.\cite{gammadetails} The two kinds alternate on the surface,
which gives a $(1\times4)$ periodicity. The spacing between two
troughs of the same kind is then 16 \AA, which corresponds to the
spacing between the NWs observed in experiment.

We use the $\gamma$-surface as a template to adsorb Pt or Ge
dimers. Of the many possible adsorption sites we have studied,
only the structure labeled E in \ref{fig1:geometries}(c) gives
rise to NWs that can match the experiment.\cite{gammadetails}
The formation energy is substantial, i.e. $E_{\mathrm{f}} =
-1.50,-1.00$ eV for Pt and Ge NWs, respectively. The large
formation energy for Ge NWs can immediately be attributed to the
formation of Pt-Ge bonds with the Pt atoms in the surface, which
is energetically advantageous. The large formation energy for Pt
NWs might be surprising at first sight, since the formation of
Pt-Pt bonds was avoided before (see the discussion above).
However, examination of the optimized structure shows that the Pt
NW has in fact sunken into the through, so that the Pt atoms of
the NW are $\sim 0.7$ \AA\ \textit{below} the average level of the
atoms in the surface top layer. In fact, these Pt atoms make bonds
with Ge atoms in the second and third layer, which explains the
stability of the structure.

The same does not happen to a Ge NW. The Ge atoms remain at a
height of $\sim 0.7$ \AA\ above the average height of atoms in the
surface top layer. The simulated STM image of a Ge NW is reasonably
close to the experimental image with bright features at the
position of the NW. However these features are not double peaked as
experimentally observed for the NWs (Fig. \ref{fig3:STMs}(f)).
Moreover, the structure of a Ge NW adsorbed on a $\gamma$-surface contains the same number of Pt atoms as the structure shown in Fig. \ref{fig3:STMs}(a). Yet its formation energy is
not as favorable, which
makes it a metastable structure.

The Pt NW has a low energy, but its simulated STM image, as shown
in Fig. \ref{fig3:STMs}(c), strongly deviates from the
experimental STM image. In fact, in the simulated image the Pt NW
is not visible at all. This is partly due to the fact that the Pt
NW has sunk into the surface, but also because Pt atoms do not
give rise to a LDOS close to the Fermi level that emerges from the
surface. The bright features in Fig. \ref{fig3:STMs}(c) correspond
to Ge atoms of the surface top layer. These Ge atoms belong to
Pt-Ge dimers that become strongly tilted after adsorption of the
Pt NW. The tilting angle of these dimers is $\sim 60^{\circ}$,
whereas the tilting angle of Pt-Ge dimers on the clean
$\gamma$-surface is $\lesssim 5^{\circ}$. The tilting is
accompanied by an increase of the Pt-Ge bond length to 2.64 \AA,
as compared to 2.35 \AA\ in the $\gamma$-surface. The Pt atoms of
the tilted Pt-Ge dimers go subsurface to form extra Pt-Ge bonds,
whereas the Ge atoms stick out of the surface and give rise to
bright features. The simulated image of the Pt NW shows in fact a
remarkable resemblance to the wide troughs observed in
Ref.~\onlinecite{Fischer:prb07}, suggesting that these features
indeed involve subsurface Pt. Note that in our structure only
every other Pt-Ge dimer along a dimer row is tilted, doubling the
periodicity along the row compared to the $\gamma$-surface to
$(2\times4)$, as is observed experimentally, see
Fig.\ref{fig3:STMs}(e).\cite{Fischer:prb07}

We have seen that it is energetically advantageous to incorporate
Pt adatoms in a trough in the $\gamma$-surface. One can therefore
imagine the following scenario. Let the Pt atoms sink into the
trough and exchange with Ge atoms in the third layer of the
substrate, as indicated schematically in Fig.
\ref{fig1:geometries}(d). We call this the $\gamma^*$-surface. The displaced Ge atoms are then pushed up from the trough and can form a NW on top with atoms in the E positions. The calculated height of these Ge NWs is $\sim 0.6$ \AA\ above the surface, which is in fair agreement with the corrugation deduced from STM line scans.\cite{fn:Fig2Zand}
All Pt atoms in this scenario form bonds with neighboring Ge
atoms, and the formation of Pt-Pt bonds is avoided altogether. The
resulting structure is energetically very favorable with a
formation energy $E_{\mathrm{f}} = -2.06$ eV. The exchange that is
required between Pt and Ge atoms in the third surface layer might
explain the high annealing temperature ($T=1050$ K) that is needed
to form the NWs experimentally. As a check we have also replaced
the Ge NW by a Pt nanowire, which leads to a substantially less
favorable formation energy $E_{\mathrm{f}} = -0.96$ eV.

The simulated STM image of a Ge NW on a $\gamma^*$-surface is shown in
Fig. \ref{fig3:STMs}(d). It is in very good agreement with the
experimental STM image of Fig. \ref{fig3:STMs}(f). All the
features of the experimental image are present in the simulated
one, including the double peak structure associated with each
dimer of the Ge NW and the bright features that are arranged
symmetrically alongside the NW. The latter result from Ge atoms
belonging to Pt-Ge dimers in the surface top layer, whereas the Pt
atoms remain ``invisible''. Replacing the Ge NW by a Pt NW
completely removes the NW in the simulated image, which clearly
indicates that the NW consists of Ge dimers. The Ge-Ge bond length
of the NW dimers is 2.72 \AA, which is somewhat larger than the
typical bond length of 2.45 \AA\ of a Ge-Ge dimer on the clean
Ge(001) surface.

In conclusion, we have studied possible structures of the Pt-Ge
surface that emerge after deposition of Pt on the Ge(001) surface.
The experimentally observed $\beta$-surface has $0.25$ ML of Pt in
its top layer, which consists of Pt-Ge and Ge-Ge dimers arranged
in a checkerboard $c(2\times4)$ pattern. Starting from either the
$\beta$-surface, or the clean Ge(001) surface, we find that Pt NWs
are unstable structures. Pt atoms have a tendency to be
incorporated in the substrate and form additional Pt-Ge bonds. We propose the
$\gamma$-structure, which contains $0.5$ ML of Pt in the surface
top layer, as a template for NWs. It consists of rows of Pt-Ge
dimers in the top layer resulting in $(1\times4)$ periodicity. The
trough between the rows lined up with Pt atoms is the most
favorable adsorption site for NWs. Adsorbing a Pt NW we observe
that it sinks into the surface and increases the width of the
trough in the STM image. Exchanging the ``sunken'' Pt atoms with
Ge atoms in the third layer of the substrate leads to a Ge NW.
This structure has a favorable formation energy and gives
simulated images in agreement with the experimental STM images.

We thank Harold Zandvliet and Arie van Houselt for stimulating discussions and for making
available their experimental STM results. This work is part of the research program of the
``Stichting voor Fundamenteel Onderzoek der Materie'' (FOM); the use of superconmputer
facilities is sponsored by the ``Stichting Nationale Computer Faciliteiten'' (NCF), both
financially supported by the ``Nederlandse Organisatie voor Wetenschappelijk Onderzoek'' (NWO).

\bibliography{nanowirenotes,cmsdanny}

\begin{thebibliography}{24}
\expandafter\ifx\csname natexlab\endcsname\relax\def\natexlab#1{#1}\fi
\expandafter\ifx\csname bibnamefont\endcsname\relax
  \def\bibnamefont#1{#1}\fi
\expandafter\ifx\csname bibfnamefont\endcsname\relax
  \def\bibfnamefont#1{#1}\fi
\expandafter\ifx\csname citenamefont\endcsname\relax
  \def\citenamefont#1{#1}\fi
\expandafter\ifx\csname url\endcsname\relax
  \def\url#1{\texttt{#1}}\fi
\expandafter\ifx\csname urlprefix\endcsname\relax\def\urlprefix{URL }\fi
\providecommand{\bibinfo}[2]{#2}
\providecommand{\eprint}[2][]{\url{#2}}

\bibitem[{\citenamefont{Barth et~al.}(2005)\citenamefont{Barth, Constantini,
  and Kern}}]{Barth:nat05}
\bibinfo{author}{\bibfnamefont{J.~V.} \bibnamefont{Barth}},
  \bibinfo{author}{\bibfnamefont{G.}~\bibnamefont{Constantini}},
  \bibnamefont{and} \bibinfo{author}{\bibfnamefont{K.}~\bibnamefont{Kern}},
  \bibinfo{journal}{Nature} \textbf{\bibinfo{volume}{437}},
  \bibinfo{pages}{671} (\bibinfo{year}{2005}), \bibinfo{note}{and references
  therein.}

\bibitem[{\citenamefont{Yeom et~al.}(1999)\citenamefont{Yeom, Takeda,
  Rotenberg, Matsuda, Horikoshi, Schaefer, Lee, Kevan, Ohta, Nagao
  et~al.}}]{Yeom:prl99}
\bibinfo{author}{\bibfnamefont{H.~W.} \bibnamefont{Yeom}},
  \bibinfo{author}{\bibfnamefont{S.}~\bibnamefont{Takeda}},
  \bibinfo{author}{\bibfnamefont{E.}~\bibnamefont{Rotenberg}},
  \bibinfo{author}{\bibfnamefont{I.}~\bibnamefont{Matsuda}},
  \bibinfo{author}{\bibfnamefont{K.}~\bibnamefont{Horikoshi}},
  \bibinfo{author}{\bibfnamefont{J.}~\bibnamefont{Schaefer}},
  \bibinfo{author}{\bibfnamefont{C.~M.} \bibnamefont{Lee}},
  \bibinfo{author}{\bibfnamefont{S.~D.} \bibnamefont{Kevan}},
  \bibinfo{author}{\bibfnamefont{T.}~\bibnamefont{Ohta}},
  \bibinfo{author}{\bibfnamefont{T.}~\bibnamefont{Nagao}},
  \bibnamefont{et~al.}, \bibinfo{journal}{Phys. Rev. Lett.}
  \textbf{\bibinfo{volume}{82}}, \bibinfo{pages}{4898} (\bibinfo{year}{1999}).

\bibitem[{\citenamefont{Lee et~al.}(2005)\citenamefont{Lee, Guo, and
  Plummer}}]{Lee:prl05}
\bibinfo{author}{\bibfnamefont{G.}~\bibnamefont{Lee}},
  \bibinfo{author}{\bibfnamefont{J.}~\bibnamefont{Guo}}, \bibnamefont{and}
  \bibinfo{author}{\bibfnamefont{E.~W.} \bibnamefont{Plummer}},
  \bibinfo{journal}{Phys. Rev. Lett.} \textbf{\bibinfo{volume}{95}},
  \bibinfo{pages}{116103} (\bibinfo{year}{2005}).

\bibitem[{\citenamefont{Ahn et~al.}(2005)\citenamefont{Ahn, Kang, Ryang, and
  Yeom}}]{Ahn:prl05}
\bibinfo{author}{\bibfnamefont{J.~R.} \bibnamefont{Ahn}},
  \bibinfo{author}{\bibfnamefont{P.~G.} \bibnamefont{Kang}},
  \bibinfo{author}{\bibfnamefont{K.~D.} \bibnamefont{Ryang}}, \bibnamefont{and}
  \bibinfo{author}{\bibfnamefont{H.~W.} \bibnamefont{Yeom}},
  \bibinfo{journal}{Phys. Rev. Lett.} \textbf{\bibinfo{volume}{95}},
  \bibinfo{pages}{196402} (\bibinfo{year}{2005}).

\bibitem[{\citenamefont{Snijders et~al.}(2006)\citenamefont{Snijders, Rogge,
  and Weitering}}]{Snijders:prl06}
\bibinfo{author}{\bibfnamefont{P.~C.} \bibnamefont{Snijders}},
  \bibinfo{author}{\bibfnamefont{S.}~\bibnamefont{Rogge}}, \bibnamefont{and}
  \bibinfo{author}{\bibfnamefont{H.~H.} \bibnamefont{Weitering}},
  \bibinfo{journal}{Phys. Rev. Lett.} \textbf{\bibinfo{volume}{96}},
  \bibinfo{pages}{076801} (\bibinfo{year}{2006}).

\bibitem[{\citenamefont{Eames et~al.}(2006)\citenamefont{Eames, Bonet, Probert,
  Tear, and Perkins}}]{Eames:prb06}
\bibinfo{author}{\bibfnamefont{C.}~\bibnamefont{Eames}},
  \bibinfo{author}{\bibfnamefont{C.}~\bibnamefont{Bonet}},
  \bibinfo{author}{\bibfnamefont{M.~I.~J.} \bibnamefont{Probert}},
  \bibinfo{author}{\bibfnamefont{S.~P.} \bibnamefont{Tear}}, \bibnamefont{and}
  \bibinfo{author}{\bibfnamefont{E.~W.} \bibnamefont{Perkins}},
  \bibinfo{journal}{Phys. Rev. B} \textbf{\bibinfo{volume}{74}},
  \bibinfo{eid}{193318} (\bibinfo{year}{2006}).

\bibitem[{\citenamefont{Lin et~al.}(1993)\citenamefont{Lin, Wan, Glueckstein,
  and Nogami}}]{LinXF:prb93}
\bibinfo{author}{\bibfnamefont{X.~F.} \bibnamefont{Lin}},
  \bibinfo{author}{\bibfnamefont{K.~J.} \bibnamefont{Wan}},
  \bibinfo{author}{\bibfnamefont{J.~C.} \bibnamefont{Glueckstein}},
  \bibnamefont{and} \bibinfo{author}{\bibfnamefont{J.}~\bibnamefont{Nogami}},
  \bibinfo{journal}{Phys. Rev. B} \textbf{\bibinfo{volume}{47}},
  \bibinfo{pages}{3671} (\bibinfo{year}{1993}).

\bibitem[{\citenamefont{Yeom et~al.}(2005)\citenamefont{Yeom, Kim, Lee, Ryang,
  and Kang}}]{Yeom:prl05}
\bibinfo{author}{\bibfnamefont{H.~W.} \bibnamefont{Yeom}},
  \bibinfo{author}{\bibfnamefont{Y.~K.} \bibnamefont{Kim}},
  \bibinfo{author}{\bibfnamefont{E.~Y.} \bibnamefont{Lee}},
  \bibinfo{author}{\bibfnamefont{K.-D.} \bibnamefont{Ryang}}, \bibnamefont{and}
  \bibinfo{author}{\bibfnamefont{P.~G.} \bibnamefont{Kang}},
  \bibinfo{journal}{Phys. Rev. Lett.} \textbf{\bibinfo{volume}{95}},
  \bibinfo{pages}{205504} (\bibinfo{year}{2005}).

\bibitem[{\citenamefont{Gurlu et~al.}(2003)\citenamefont{Gurlu, Adam,
  Zandvliet, and Poelsema}}]{Gurlu:apl03}
\bibinfo{author}{\bibfnamefont{O.}~\bibnamefont{Gurlu}},
  \bibinfo{author}{\bibfnamefont{O.~A.~O.} \bibnamefont{Adam}},
  \bibinfo{author}{\bibfnamefont{H.~J.~W.} \bibnamefont{Zandvliet}},
  \bibnamefont{and} \bibinfo{author}{\bibfnamefont{B.}~\bibnamefont{Poelsema}},
  \bibinfo{journal}{Appl. Phys. Lett.} \textbf{\bibinfo{volume}{83}},
  \bibinfo{pages}{4610} (\bibinfo{year}{2003}).

\bibitem[{\citenamefont{Schafer et~al.}(2006)\citenamefont{Schafer, Schrupp,
  Preisinger, and Claessen}}]{Schafer:prb06}
\bibinfo{author}{\bibfnamefont{J.}~\bibnamefont{Schafer}},
  \bibinfo{author}{\bibfnamefont{D.}~\bibnamefont{Schrupp}},
  \bibinfo{author}{\bibfnamefont{M.}~\bibnamefont{Preisinger}},
  \bibnamefont{and} \bibinfo{author}{\bibfnamefont{R.}~\bibnamefont{Claessen}},
  \bibinfo{journal}{Phys. Rev. B} \textbf{\bibinfo{volume}{74}},
  \bibinfo{pages}{041404(R)} (\bibinfo{year}{2006}).

\bibitem[{\citenamefont{Oncel et~al.}(2005)\citenamefont{Oncel, van Houselt,
  Huijben, Hallback, Gurlu, Zandvliet, and Poelsema}}]{Oncel:prl05}
\bibinfo{author}{\bibfnamefont{N.}~\bibnamefont{Oncel}},
  \bibinfo{author}{\bibfnamefont{A.}~\bibnamefont{van Houselt}},
  \bibinfo{author}{\bibfnamefont{J.}~\bibnamefont{Huijben}},
  \bibinfo{author}{\bibfnamefont{A.~S.} \bibnamefont{Hallback}},
  \bibinfo{author}{\bibfnamefont{O.}~\bibnamefont{Gurlu}},
  \bibinfo{author}{\bibfnamefont{H.~J.~W.} \bibnamefont{Zandvliet}},
  \bibnamefont{and} \bibinfo{author}{\bibfnamefont{B.}~\bibnamefont{Poelsema}},
  \bibinfo{journal}{Phys. Rev. Lett.} \textbf{\bibinfo{volume}{95}},
  \bibinfo{pages}{116801} (\bibinfo{year}{2005}).

\bibitem[{\citenamefont{van Houselt et~al.}(2006)\citenamefont{van Houselt,
  Oncel, Poelsema, and Zandvliet}}]{Houselt:nanol06}
\bibinfo{author}{\bibfnamefont{A.}~\bibnamefont{van Houselt}},
  \bibinfo{author}{\bibfnamefont{N.}~\bibnamefont{Oncel}},
  \bibinfo{author}{\bibfnamefont{B.}~\bibnamefont{Poelsema}}, \bibnamefont{and}
  \bibinfo{author}{\bibfnamefont{H.}~\bibnamefont{Zandvliet}},
  \bibinfo{journal}{Nano Letters} \textbf{\bibinfo{volume}{6}},
  \bibinfo{pages}{1439} (\bibinfo{year}{2006}).

\bibitem[{\citenamefont{Fischer et~al.}(2007)\citenamefont{Fischer, van
  Houselt, Kockmann, Poelsema, and Zandvliet}}]{Fischer:prb07}
\bibinfo{author}{\bibfnamefont{M.}~\bibnamefont{Fischer}},
  \bibinfo{author}{\bibfnamefont{A.}~\bibnamefont{van Houselt}},
  \bibinfo{author}{\bibfnamefont{D.}~\bibnamefont{Kockmann}},
  \bibinfo{author}{\bibfnamefont{B.}~\bibnamefont{Poelsema}}, \bibnamefont{and}
  \bibinfo{author}{\bibfnamefont{H.~J.~W.} \bibnamefont{Zandvliet}},
  \bibinfo{journal}{Phys. Rev. B} \textbf{\bibinfo{volume}{76}},
  \bibinfo{pages}{245429} (\bibinfo{year}{2007}).

\bibitem[{\citenamefont{Gurlu et~al.}(2004)\citenamefont{Gurlu, Zandvliet,
  Poelsema, Dag, and Ciraci}}]{Gurlu:prb04}
\bibinfo{author}{\bibfnamefont{O.}~\bibnamefont{Gurlu}},
  \bibinfo{author}{\bibfnamefont{H.~J.~W.} \bibnamefont{Zandvliet}},
  \bibinfo{author}{\bibfnamefont{B.}~\bibnamefont{Poelsema}},
  \bibinfo{author}{\bibfnamefont{S.}~\bibnamefont{Dag}}, \bibnamefont{and}
  \bibinfo{author}{\bibfnamefont{S.}~\bibnamefont{Ciraci}},
  \bibinfo{journal}{Phys. Rev. B} \textbf{\bibinfo{volume}{70}},
  \bibinfo{pages}{085312} (\bibinfo{year}{2004}).

\bibitem[{com()}]{compdetails}
\bibinfo{note}{We use a plane wave basis set and the PAW formalism
  \cite{Bloechl:prb94,Kresse:prb99} at the level of the local density
  approximation, as implemented in the VASP
  code.\cite{Kresse:prb93,Kresse:prb96} The plane wave kinetic energy cutoff is
  set at 345 eV. The supercell contains a symmetric slab of 12 layers of Ge
  atoms. Pt atoms are added or replace Ge atoms on both (top and bottom)
  surfaces. A vacuum region of $\sim 15.5$ \AA\ separates the periodic images
  of the slab. We use a $8\times4$ \textbf{k}-point grid to sample the
  Brillouin zone of the $(2\times4)\mathrm{R45^o}$ surface unit cell. To
  optimize the geometry we apply the conjugate gradient algorithm while keeping
  the positions of the germanium atoms in the center two layers fixed at their
  bulk positions.}

\bibitem[{sim()}]{simudetails}
\bibinfo{note}{The Tersoff-Hamann model is used, in which tunneling currents
  are represented by integrating the local density of states (LDOS) of the
  surface over a range that corresponds to the applied bias
  $V$.\cite{Tersoff:prb85} We mimic the STM constant current mode by tracing a
  constant integrated LDOS $\rho(x,y,z)=\rho_c$ as a function of $x,y$ and
  mapping $z$ on a grayscale. At the position of the highest atom $z=3.0$ \AA.}

\bibitem[{fac()}]{factornote}
\bibinfo{note}{Dividing the number by two, since our slabs contain two
  identical surfaces at the top and bottom.}

\bibitem[{\citenamefont{Zandvliet}()}]{fn:Fig2Zand}
\bibinfo{author}{\bibfnamefont{H.~J.~W.} \bibnamefont{Zandvliet}},
  \bibinfo{note}{unpublished.}

\bibitem[{gam()}]{gammadetails}
\bibinfo{note}{We have also studied other structures corresponding to 0.5ML of
  Pt in the surface and Pt or Ge nanowires adsorbed at different positions.
  None of these lead to strucures that resemble the experimental STM images.}

\bibitem[{\citenamefont{Bl{\"{o}}chl}(1994)}]{Bloechl:prb94}
\bibinfo{author}{\bibfnamefont{P.~E.} \bibnamefont{Bl{\"{o}}chl}},
  \bibinfo{journal}{Phys. Rev. B} \textbf{\bibinfo{volume}{50}},
  \bibinfo{pages}{17953} (\bibinfo{year}{1994}).

\bibitem[{\citenamefont{Kresse and Joubert}(1999)}]{Kresse:prb99}
\bibinfo{author}{\bibfnamefont{G.}~\bibnamefont{Kresse}} \bibnamefont{and}
  \bibinfo{author}{\bibfnamefont{D.}~\bibnamefont{Joubert}},
  \bibinfo{journal}{Phys. Rev. B} \textbf{\bibinfo{volume}{59}},
  \bibinfo{pages}{1758} (\bibinfo{year}{1999}).

\bibitem[{\citenamefont{Kresse and Hafner}(1993)}]{Kresse:prb93}
\bibinfo{author}{\bibfnamefont{G.}~\bibnamefont{Kresse}} \bibnamefont{and}
  \bibinfo{author}{\bibfnamefont{J.}~\bibnamefont{Hafner}},
  \bibinfo{journal}{Phys. Rev. B} \textbf{\bibinfo{volume}{47}},
  \bibinfo{pages}{(R)558} (\bibinfo{year}{1993}).

\bibitem[{\citenamefont{Kresse and Furthm{\"{u}}ller}(1996)}]{Kresse:prb96}
\bibinfo{author}{\bibfnamefont{G.}~\bibnamefont{Kresse}} \bibnamefont{and}
  \bibinfo{author}{\bibfnamefont{J.}~\bibnamefont{Furthm{\"{u}}ller}},
  \bibinfo{journal}{Phys. Rev. B} \textbf{\bibinfo{volume}{54}},
  \bibinfo{pages}{11169} (\bibinfo{year}{1996}).

\bibitem[{\citenamefont{Tersoff and Hamann}(1985)}]{Tersoff:prb85}
\bibinfo{author}{\bibfnamefont{J.}~\bibnamefont{Tersoff}} \bibnamefont{and}
  \bibinfo{author}{\bibfnamefont{D.~R.} \bibnamefont{Hamann}},
  \bibinfo{journal}{Phys. Rev. B} \textbf{\bibinfo{volume}{31}},
  \bibinfo{pages}{805} (\bibinfo{year}{1985}).

\end{thebibliography}

\end{document}